**Phase fluctuations dominated superconductivity in a thick Granular Boron-Doped Diamond film**

Sulagna Dutta[1], Subhajit Chatterjee[2,3], Vivas Bagwe[1], Pritam Das[1], S. Anantha Bharathi[4], C. Nikhil[3,4], M.S. Ramachandra Rao[2,3,4*] and Pratap Raychaudhuri[1†]

*[1] Tata Institute of Fundamental Research, Homi Bhabha Road, Mumbai 400005.*
*[2] Department of Physics, Indian Institute of Technology (IIT) Madras, Chennai 600036.*
*[3] Quantum Centre of Excellence for Diamond and Emergent Materials (QuCenDiEM), Indian Institute of Technology (IIT) Madras, Chennai – 600036.*
*[4] India Centre for Lab-grown Diamond (InCent-LGD), Indian Institute of Technology (IIT) Madras, Chennai 600036.*

We investigate the superconducting properties of a 1.2-μm-thick granular boron-doped diamond film, where tightly packed grains form a disordered network of Josephson junctions. The film exhibits a sharp superconducting transition at a critical temperature, $T_c \sim 3$ K. Its transport properties are well described by the Ambegaokar-Halperin theory of phase fluctuations. Direct measurements of the penetration depth using a two-coil mutual inductance technique reveal classic phase-fluctuation-dominated behavior: below $T_c$, the superfluid density ($n_s$) varies linearly over a wide temperature range—a hallmark of thermal phase fluctuations—before crossing over to a $T^2$ dependence at ultra-low temperatures. Both the large low-temperature penetration depth, $\lambda(0) \approx 8.8$ μm, and the slope of the linear $n_s$ variation in the thermal regime reveal a remarkably low superfluid stiffness in the range of 0.5–0.8 K. These results establish granular boron-doped diamond films as an excellent model system for investigating the role of long-wavelength phase fluctuations in three-dimensional superconductors.

* E-mail: msrrao@physics.iitm.ac.in
† E-mail: pratap@tifr.res.in

At a fundamental level, superconductivity is governed by two energy scales: the superconducting energy gap, $\Delta$, which is the energy cost of exciting quasiparticles by breaking a Cooper pair, and the superfluid stiffness, $J_s$, which is the energy associated with generating a phase twist in the phase-coherent Cooper pair condensate[1,2]. In conventional 3-dimensional superconductors, $J_s \gg \Delta$, meaning the temperature dependence of the superconducting properties are governed by $\Delta$ alone. This premise underpins Bardeen-Cooper-Schrieffer (BCS) theory (or its extensions) where the superconducting transition temperature, $T_c$, scales proportionally with $\Delta$, yielding the celebrated universal BCS ratio[3], $\frac{\Delta}{k_B T_c} \sim 1.76$ ($k_B$ is the Boltzmann constant). On the other hand, there are certain situations, such as disordered or low dimensional superconductors, where $J_s$ can become smaller than $\Delta$. In such scenarios, phase fluctuations can dismantle superconductivity long before the pairing interaction itself vanishes[2,4], such that $k_B T_c = A J_s$ ,where $A$ is a constant of the order of unity. Recent discoveries of various correlated and two-dimensional systems—where phase fluctuations are thought to play a dominant role—have sparked significant interest in how these dynamics alter the superconducting state[5,6,7,8].

The paradigmatic example of phase fluctuation in superconductors is the Berezinskii-Kosterlitz-Thouless transition[9,10,11], typically realized in thin films where the thickness ($t$) is less than the coherence length ($\xi$). In these 2-dimensional systems, vortical excitations of the phase within the Cooper-pair condensate dictate the unbinding of vortex-antivortex pairs. This topological mechanism drives the superfluid density ($n_s$) abruptly to zero[12,13,14,15] at a critical temperature significantly suppressed relative to the predicted mean-field $T_c$. Conversely, the impact of phase fluctuations in three-dimensional (3D) superconductors has remained comparatively less studied. In 3D superconductors, the formation of topological defects like vortices is energetically unfavorable; consequently, the excited phase modes are characteristically long-wavelength in nature. While simultaneously modeling the combined effects of quasiparticle excitations and long-wavelength phase fluctuations (LWPF) presents a formidable theoretical challenge, definitive predictions can be extracted in the asymptotic limit where $J_s \ll \Delta$, from the phase only Hamiltonian[1,2],

$$H = \frac{1}{2}\rho_s \int d^3r \, |\nabla\theta|^2 \ , \quad \rho_s = \frac{\hbar^2 n_s^0}{4m^*} \quad (1)$$

where, $n_s^0$ is the superfluid density in the absence of phase fluctuations, $m^*$ and $e$ are the effective mass and charge of the electron, $\mu_0$ is the vacuum permittivity, $\hbar$ is the reduced Plank's constant and $\nabla\theta$ is the phase-gradient in the superconducting condensate. In 3D, $\rho_s$ has the dimension of energy-(length)$^{-1}$. Consequently, the energy scale $J_s$ is defined by multiplying $\rho_s$ with a short wavelength cutoff $a$ such that,

$$J_s = \rho_s a \approx \frac{\hbar^2 a}{4\mu_0 \lambda(0)^2 e^2}. \quad (2)$$

The last form in eqn. (2) (strictly valid in the absence of quantum phase fluctuations) is obtained by connecting $n_s^0$ to the low temperature penetration depth $\lambda(0)$ using the London relation[3], $n_s = \frac{m^*}{\mu_0 \lambda^2 e^2}$. Within the spin-wave approximation, $n_s$ depends on the root mean square amplitude of phase fluctuations $\langle(\Delta\theta)^2\rangle$ as[16,17,18],

$$n_s = n_s^0 \left(1 - \frac{\langle(\Delta\theta)^2\rangle}{2D}\right) \qquad (3)$$

where $\langle(\Delta\theta)^2\rangle$ is the root-mean-square amplitude of phase fluctuations, and $D$ is the dimension ($D$ = 3 in our case). Within mean-field approximation, the Hamiltonian in eqn .(1) yields three explicit consequences. For thermal fluctuations[2], $\langle(\Delta\theta)^2\rangle = \frac{k_B T}{J_s}$. Thus $n_s$ linearly increases below the superconducting transition temperature. At low temperatures, where thermal phase fluctuations become small, the linear increase gets limited to a value determined by quantum phase fluctuations, for which[2] $\langle(\Delta\theta)^2\rangle = \frac{1}{2}\sqrt{\frac{E_c}{J_s}}$, where $E_c$ is the Coulomb charging energy. Consequently, the cross-over from thermal to quantum behavior is expected at a characteristic temperature, $T^{cross}$ corresponding Josephson plasma frequency, $k_B T^{cross} \approx \hbar\omega_p \sim \sqrt{E_c J_s}$. To date, empirical validation of this behavior has been observed in highly disordered s-wave superconductors[5,6,8], where strong disorder scattering drives the system into the phase-fluctuation-dominated regime. However, additional complications in those systems, such as the presence of emergent inhomogeneity, complicates the analysis and prevents a quantitative comparison with theory. On the other hand, granular superconductors, where individual superconducting grains are connected to each other through weak links, is another class of systems where phase fluctuations can play a significant role[19]. Here, the sample can be visualized as a disordered network of Josephson junctions (JJ) where the Josephson coupling energy between grains, $E_J = \frac{\hbar I_c^0}{2e}$ (where, $I_c^0$ is the critical current of the junction and $e$ is the electron charge) determines the bulk $J_s$. In such a system, one can expect phase fluctuations to dominate the superconducting properties if $E_J$, and hence $J_s$, is sufficiently low. The Josephson junction network (JJN) description has been invoked to explain the transport properties of quenched-condensed granular superconducting films[20] and granular High-$T_c$ cuprates[21,22]. However, a detailed study of phase fluctuations in granular superconductors, from which one can test the theoretical predictions of LWPF has not been carried out.

In this paper, we study granular thin films of boron-doped diamond (BDD), a conventional superconductor with low carrier density[23,24,25]. We show that the current voltage (*I-V*) characteristics of the sample in the superconducting state are well described by Ambegaokar-Halperin (AH) theory[26] of thermally activated phase slips. Furthermore, we directly measure the penetration depth, $\lambda$, and show that the temperature variation of $n_s$ follows the predictions of LWPF both for thermal and quantum phase fluctuations, thus giving a near perfect realization of phase-fluctuation dominated superconductivity.

The BDD film used in this study was grown through hot filament chemical vapor deposition[27,28] on a silicon substrate. Top-view scanning electron microscopy (SEM) images (Fig. 1(a)) reveal that the film consists mainly of densely packed crystallites with sizes varying from 50-200 nm, whereas a few large grains with sizes up to 500 nm across their widest lateral dimensions can also be seen. The cross-sectional SEM view (Fig. 1(b)) shows that these crystallites are elongated and imperfectly aligned perpendicular to the growth direction. From Hall-effect measurement, we obtain a carrier concentration of $2.18 \times 10^{27}\ m^{-3}$. Fig. 1(c) shows the resistance versus temperature (*R-T*) measured at various fields, on a sample patterned in a Hall bar geometry; the corresponding resistivity, $\rho$, is shown on the right axis. In zero field, we observe that the resistance starts gradually decreasing below 13 K followed by a sharp transition to a zero-resistance state at $T_{c0}{\sim}3$ K. An expanded view of the *R-T* curves (inset, Fig. 1(d)) shows that the resistance in different fields starts diverging much above $T_{c0}$, suggesting that the onset of the superconducting transition happens at a much higher temperature. From the temperature where $\frac{\rho(60\ kOe)-\rho(0)}{\rho(0)} \approx 0.005$ we define the onset temperature, $T_c^{on} \sim 12.3$ K. These observations are similar to ref. 29 where, in addition, it was observed from photoemission spectroscopy measurements that the superconducting energy gap closes at the onset temperature. Connecting these observations, we interpret $T_c^{on}$ as the temperature where superconductivity starts nucleating in individual grains, whereas the final transition to the zero-resistance state at $T_{c0}$ marks the onset of global phase coherence. At 60 kOe, where the film is in the normal state, $\frac{d\rho}{dT} < 0$ down to 300 mK. In Fig. 1(d) we plot the conductivity, $\sigma = 1/\rho$, as a function of temperature. Below 3 K, the temperature variation of $\sigma$ follows the functional form, $\sigma = \sigma_0 + AT^{1/2}$ as theoretically predicted[30] for a granular metal at low temperatures.

To confirm that the film behaves like a JJN, we first analyze the current-voltage (*I-V*) characteristics measured at various temperatures in zero field (Fig. 2(a)). To reduce Joule heating, the measurement was done in pulsed mode with current pulse duration of 50 ms. The *I-V* curves display a prominent rounding off at the onset of voltage followed by a sharp transition to normal state. This rounding off, caused by thermally activated phase slips is characteristic of a Josephson junction. To quantitatively analyze the *I-V* curve before the abrupt jump, we use the Ambegaokar-Halperin (AH) model[26,3], where the *I-V* curve for a single JJ is expressed in terms of the thermal noise rounding parameter, $\gamma = \frac{2E_J}{k_B T}$, as,

$$V = I_c R_N \frac{4\pi}{\gamma}\left[\left\{\frac{1}{(e^{\pi\gamma x}-1)}\right\}\left\{\int_0^{2\pi} d\theta f(\theta)\right\}\left\{\int_0^{2\pi} d\theta' \frac{1}{f(\theta')}\right\} + \int_0^{2\pi} d\theta \int_\theta^{2\pi} d\theta' \frac{f(\theta)}{f(\theta')}\right]^{-1} \quad (4)$$

where, $x = I/I_c$, $f(\theta) = e^{-U(\theta)/k_B T}$, $U(\theta) = -\frac{1}{2}\gamma k_B T(x\theta + \cos\theta)$ and $R_N$ is the normal state resistance. In a JJN, the total current is divided through several parallel superconducting channels, *N*, such that $I_c = NI_c^0$, where $I_c^0$ is the average critical current for a single JJ. Using $\gamma$ and $I_c$ as independent fitting parameters, Eqn. (4) captures the low voltage rounding of the *I-V* curves as shown in Fig. 2(b). The best fit values of $\gamma$ and $I_c$ are shown in Fig. 2(c). In the graph we also plot $I_c'$, the critical current defined using a fixed voltage criterion of 100 μV. We note that $I_c' < I_c$; this is expected since the second effect of thermal phase fluctuations is to decrease the threshold current at which finite voltage appears. From $\gamma$ we obtain $\frac{E_J}{k_B} \sim 31$ K at 450 mK, which is an effective value of $E_J/k_B$ averaged over all parallel channels. However, this value is not truly reflective of $J_s$ for a granular sample where one can expect a large distribution of $E_J$ across individual grains. Since transport current preferentially flows through the path of least resistance, close to the onset of the voltage threshold the *I-V* characteristics are strongly dominated by channels with the largest critical currents. Consequently, $\gamma$ extracted from the AH fit will be strongly dominated by the links with largest Josephson coupling energies. The actual value of $J_s$ is in fact much smaller as we will show below.

We now focus on the temperature variation of $n_s$ which is the central result of this paper. We measure $\lambda$ using a two-coil mutual inductance technique[31,32,2,33]. In this technique a 8 mm diameter BDD film (obtained from the same growth run) is sandwiched between a quadrupolar primary coil and a dipolar secondary coil (*inset* Fig. 3(a)), and the magnetic shielding response of the film is measured from the real and imaginary component of the mutual inductance, $M = M' + iM''$. To avoid any non-linear effect the a.c. excitation field is kept very low ( ~ 1.6 mOe) with a frequency of 30 kHz. In Fig. 3(a) we show the temperature variation of $M'$ and $M''$. $M'$ drops sharply at $T_{c0}$ signaling the onset of magnetic shielding. At finite frequencies the response of a superconductor is described by a complex penetration depth, $\tilde{\lambda}^{-2} = \lambda^{-2} + i\delta^{-2}$, where $\delta$ is the skin depth associated with the dissipative response arising from normal electrons. To obtain the temperature variation of $\lambda$, we invert the *M-T* data by simultaneously solving the coupled Maxwell and London equations and determine $\lambda$ (and $\delta$) at each temperature. Details of this inversion procedure have been described elsewhere[31,34]. At 300 mK, $\lambda \sim 8.8\ \mu$m. Such uncharacteristic large value of $\lambda$ for a conventional superconductor, has earlier been reported for BDD films from microwave measurements[35]. Fig. 3(b) shows the temperature variation of $\lambda^{-2}$ and $\delta^{-2}$. $\lambda^{-2}$ increases monotonically below $T_{c0}$, whereas $\delta^{-2}$ shows a characteristic dissipative peak close to the superconducting transition. In Fig. 3(c) we plot the temperature variation of $n_s = \frac{m^*}{\mu_0 \lambda^2 e^2}$ (Fig. 3(c)). The band averaged effective mass in BDD obtained from different measurements varies in the range[36,37,38] $m^* = (0.75 - 1.3)\ m_e$, where $m_e$ is the free electron mass. Since the precise value of $n_s$ within a factor of unity does not affect our analysis, we assume, $m^* = m_e$. There are two central observations: (i) $n_s$ increases linearly below $T_{c0}$ down to ~1 K and then (ii) $n_s$ tends to saturate towards a limiting value below 1 K.

To quantitatively analyze these data, we first focus on the thermal phase fluctuation regime where $n_s$ varies linearly with temperature. To estimate $J_s$, we fit $n_s - T$ data above 1.2 K with eqn. (3) and $\langle(\Delta\theta)^2\rangle = \frac{k_B T}{J_s}$ (Fig. 3(c)), we obtain $J_s/k_B \sim 0.5$ K and $n_s^0 = 4.8 \times 10^{23}\ m^{-3}$. Using the value of $\lambda$ at 300 mK, from Eqn. (2) this gives, $a \approx 6.1$ nm. This is of the same order of magnitude as the coherence length $\xi = 8.9\ nm$, obtained from the upper critical field[3] (see supplementary material) at 300 mK, $H_{c2} = \frac{\Phi_0}{2\pi\xi^2} \sim 41\ kOe$, where $\Phi_0 = \frac{h}{2e}$ is the flux quantum. Below 1.1 K, we observe that $n_s$ follows almost perfectly a variation of the form $n_s = n_s^{q0}(1 - BT^2)$. The $T^2$ at low temperatures has been predicted from quantum fluctuations, when there are low energy sub-gap dissipative modes in the superconductor[16,39]. In disordered JJNs the existence of such low energy modes has been speculated from numerical simulations[40]. Consequently, $n_s^{q0}$ is the quantum-fluctuations limited value of the superfluid density in the zero-temperature limit. Therefore, the extent of suppression of superfluid density from quantum fluctuations is obtained from the ratio $\frac{n_s^{q0}}{n_s^0} \sim 0.77$. Theoretically, from eqn. (3), $\frac{n_s^{q0}}{n_s^0} = \left(1 - \frac{1}{12}\sqrt{\frac{E_c}{J_s}}\right)$. Since an independent estimate of $E_c$ is difficult for a granular sample with a broad distribution of grain size, we investigate the consistency between the quantum to classical crossover temperature, $T^{cross} \approx \frac{\sqrt{E_c J_s}}{k_B}$ with $\frac{n_s^{q0}}{n_s^0}$. Using $T^{cross} \sim 1.1$ K and $J_s/k_B \sim 0.5$ K, we obtain $E_c/k_B = 2.42$ K , giving $\frac{n_s^{q0}}{n_s^0} = 0.82$. This is remarkably close to the experimental value. The slightly larger value of the theoretical estimate is in fact consistent with the fact that low energy dissipative modes which give rise to the $T^2$ dependence of $n_s$, also makes quantum phase fluctuations somewhat less effective in suppressing the superfluid density.

In summary, we have validated the central predictions of LWPF from a direct measurement of $n_s$ from the penetration depth, in a 3-dimensional granular BDD sample. The key ingredient in our sample that allows this neat manifestation of phase fluctuations is the weak Josephson coupling between grains that causes the large separation between $T_{c0}$ and $T_c^{on}$ and ensures that the temperature evolution of the $n_s$ is dominated by phase fluctuations. However, this is not true in all granular samples. For example, when the intergranular Josephson coupling is strong enough, quasiparticle excitations and phase fluctuations would operate on comparable footing. In fact, in some of the BDD films from other growth runs with higher $T_{c0}$ were in this regime where $n_s$ followed a more complex behavior (See supplementary document). How to address those situations currently remains an outstanding challenge that needs to be investigated in future studies.

Acknowledgements: Authors from TIFR, Mumbai would like to acknowledge the funding support provided by the Department of Atomic Energy, Govt of India. The authors from IIT Madras would like to acknowledge the funding support provided by the Department of Science and Technology (DST), New Delhi, which facilitated the establishment of the *Nano Functional Materials Technology Centre (NFMTC)* (Grant Nos. SRNM/NAT/02-2005 and SR/NM/NT-01/2016); the Ministry of Human Resource Development (MHRD) for establishing *QuCenDiEM* under the Institution of Eminence/Centre of Excellence (IoE/CoE) initiative (Grant No. 11/9/2019-U.3(A)); and the Ministry of Commerce and Industry (MoCI) for providing funding support for the establishment of the *India Centre for Lab Grown Diamond (InCent-LGD).*

---

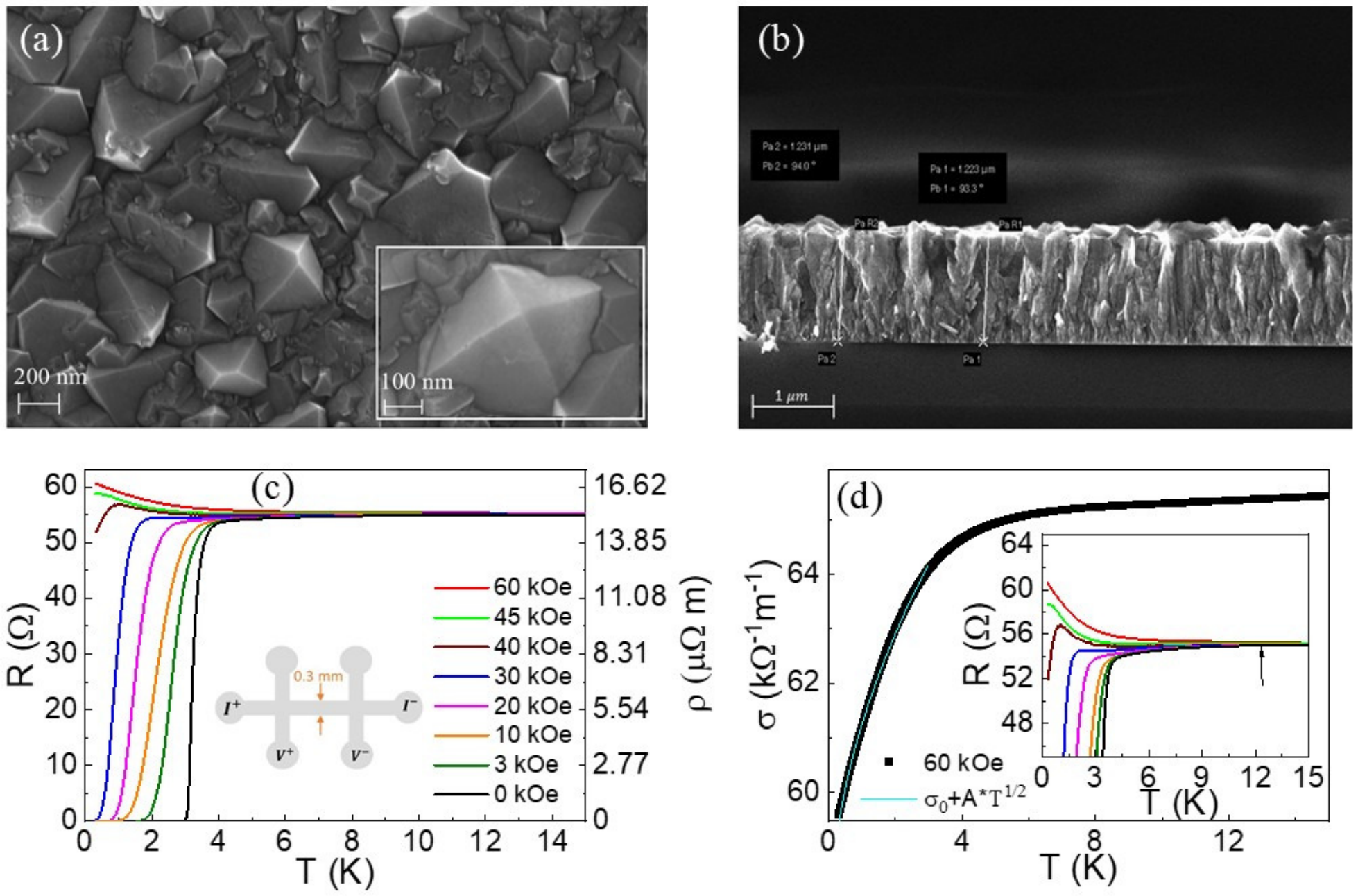

Figure 1. (a) Top-view scanning electron microscopy (SEM) image showing large crystallites (300–500 nm) embedded in a matrix of smaller crystallites. (b) Cross-sectional SEM image showing elongated crystallites, with a film thickness of approximately 1.2 $\mu m$. (c) Resistance versus temperature measured at different magnetic fields for a Hall-bar-patterned sample (*inset*); the corresponding resistivity is shown on the right axis. (d) Conductivity, $\sigma$, as a function of temperature at 60 kOe. Below 3 K, the temperature dependence follows $\sigma = \sigma_0 + AT^{1/2}$ (light blue line). The inset shows an expanded view of the $R$-$T$ curves. Below $T_c^{on} \sim 12.3$ K, the resistance curves measured at different fields begin to separate. We define $T_c^{on}$ by the temperature where $\frac{\rho(60\ kOe)-\rho(0)}{\rho(0)} \approx 0.005$. Upon further cooling, global superconductivity is established at $T_{c0} \sim 3.01$ K in zero field.

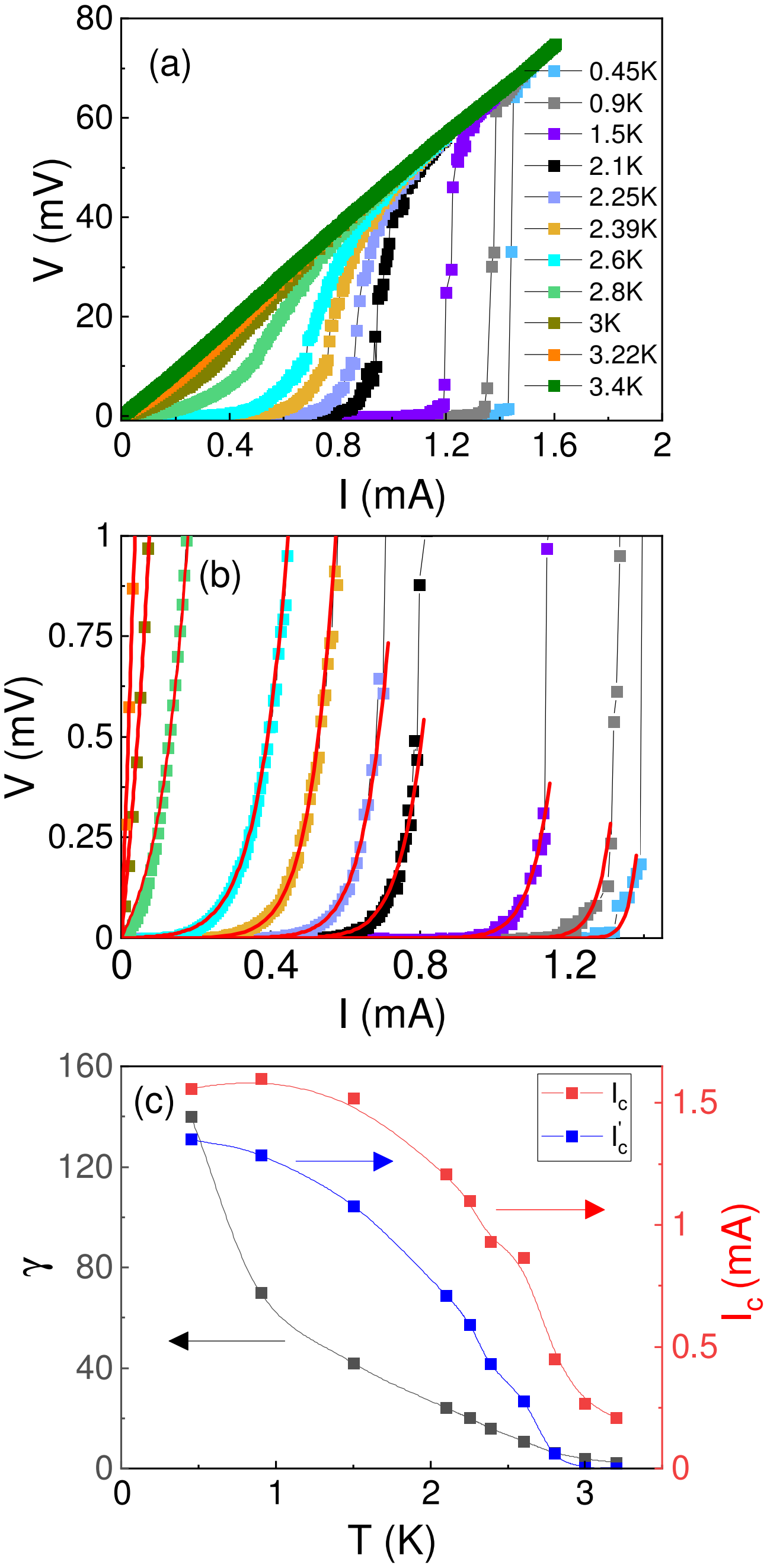

Figure 2. (a) Zero-field *I-V* characteristics measured at different temperatures. (b) Expanded view of the *I-V* curves along with the Ambegaokar-Halperin fit (eqn. 4) in the low voltage region. $\gamma = \frac{2E_J}{k_B T}$ and $I_c$ are the fitting parameters. (c) Temperature dependence of $\gamma$ and $I_c$. Also shown is $I_c'$, obtained using a fixed voltage criterion of 100 $\mu V$.

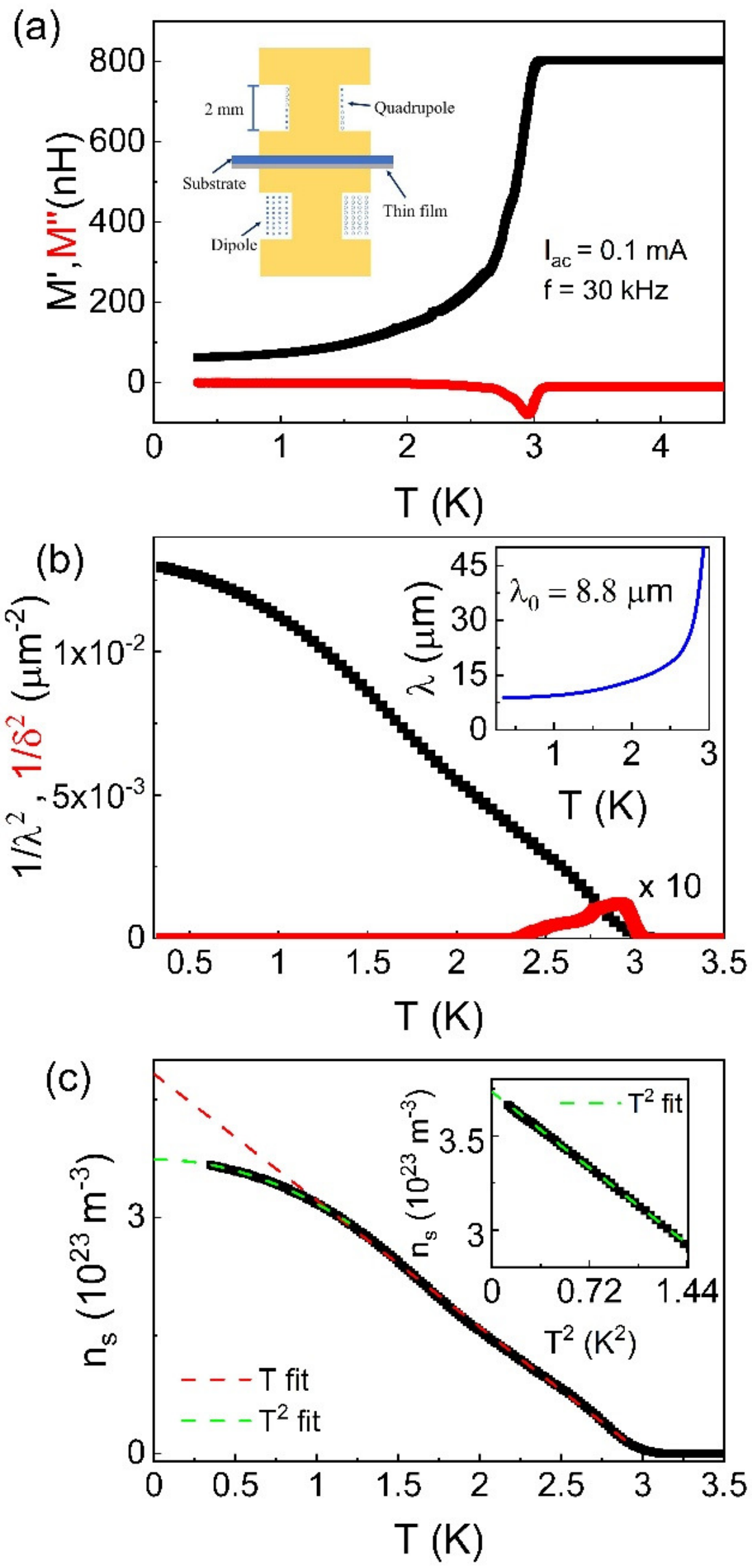

Figure 3. (a) Temperature dependence of the real and imaginary components of the mutual inductance, (M') and (M''), measured at 30 kHz with an a.c. excitation current of 0.1 mA. (*inset*) The schematics of the two-coil mutual inductance setup with the film placed between the primary and secondary coils. (b) Temperature dependence of $\lambda^{-2}$ and $\delta^{-2}$, obtained from the complex mutual inductance. The inset shows $\lambda$ variation with temperature, and $\lambda_0 = 8.8\,\mu m$. (c) Temperature dependence of the superfluid density $n_s = \frac{m^*}{\mu_0 \lambda^2 e^2}$. The dashed red and green lines denote the linear-in-$T$ and $T^2$ fits, respectively. The linear regime is consistent with thermal phase fluctuations, whereas the low-temperature $T^2$ dependence reflects the quantum-fluctuation regime in the presence of sub-gap dissipative modes. The inset shows $n_s$ plotted against $T^2$ highlighting the low-temperature quadratic dependence.

**Phase fluctuations dominated superconductivity in a thick Granular Boron-Doped Diamond film**

Sulagna Dutta[1], Subhajit Chatterjee[2,3], Vivas Bagwe[1], Pritam Das[1], S. Anantha Bharathi[4], C. Nikhil[3,4], M.S. Ramachandra Rao[2,3,41] and Pratap Raychaudhuri[12]

[1] *Tata Institute of Fundamental Research, Homi Bhabha Road, Mumbai 400005.*
[2] *Department of Physics, Indian Institute of Technology (IIT) Madras, Chennai 600036.*
[3] *Quantum Centre of Excellence for Diamond and Emergent Materials (QuCenDiEM), Indian Institute of Technology (IIT) Madras, Chennai – 600036.*
[4] *India Centre for Lab-grown Diamond (InCent-LGD), Indian Institute of Technology (IIT) Madras, Chennai 600036.*

**Supplementary Material**

**Determination of the upper critical field**

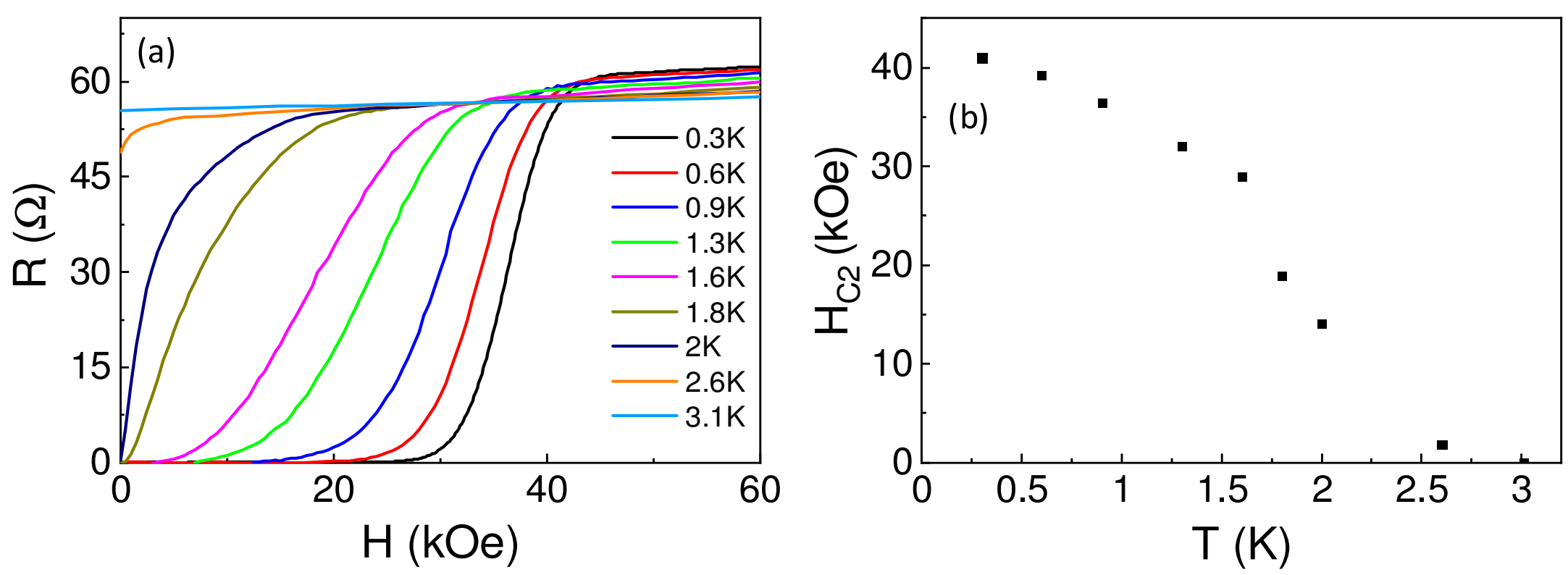

**Figure S1.** (a) Magnetic-field dependence of the resistance, $R$, measured at different temperatures. (b) Temperature dependence of $H_{C2}$. At the lowest measured temperature, $H_{C2} = 41\ kOe$.

Figure S1(a) shows the field dependence of the resistance, $R(H)$, measured at different temperatures. The upper critical field $H_{C2}$ was determined using a 90% resistance criterion. For each temperature, the resistance measured at the highest applied field of 60 $kOe$ was taken as the normal-state resistance $R_N(T) = R(60\ kOe, T)$. The normal-state resistance exhibits a negative temperature coefficient, $\frac{dR_N}{dT} < 0$, and therefore decreases with increasing temperature. Consequently, $R_N$ is temperature dependent, and the 90% criterion was applied separately to the corresponding $R_N$ at each temperature. We then define $H_{C2}(T)$ as the field at which the resistance reaches 90 % of this value,

$$R(H_{C2}, T) = 0.9 R_N(T)$$

At $T = 0.3\ K$, this criterion gives $H_{C2} = 41\ kOe$. Repeating the same procedure for each temperature yields the temperature dependence of $H_{C2}$ shown in Fig. S1(b). The extracted $H_{C2}$ decreases monotonically with increasing temperature and approaches zero near $T_{c0}$.

**Superfluid density in BDD films from different growth batches**

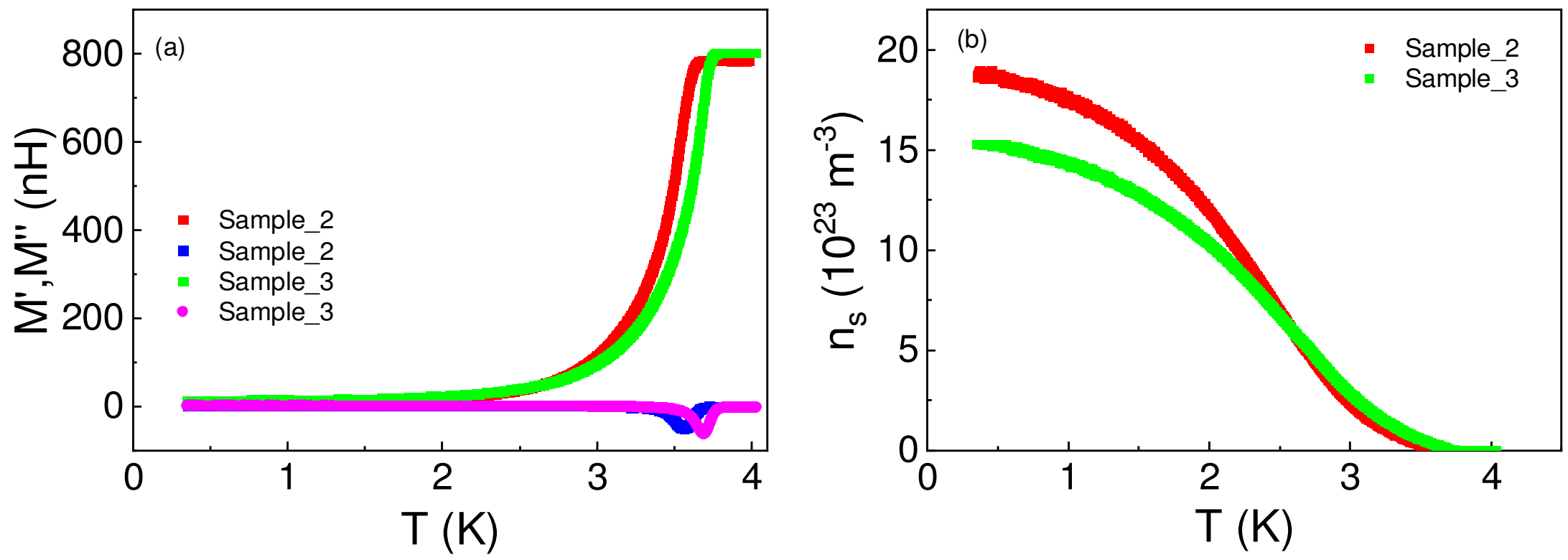

**Figure S2. (a) Temperature dependence of the real $M'$ and imaginary $M"$, components of the mutual inductance for Samples 2 and 3. (b) Corresponding superfluid density $n_s$, obtained from the penetration depth $n_s = m/(\mu_0 e^2 \lambda^2)$.**

To examine the reproducibility of the superfluid response, we performed mutual-inductance measurements on BDD films obtained from different growth batches. Figure S2(a) shows the real and imaginary components of the mutual inductance, $M'$ and $M"$, for two representative films, while Fig. S2(b) shows the corresponding temperature dependence of the superfluid density.

These films exhibit higher superconducting transition temperatures, with $T_c \sim 3.6$ K. for Sample 2 and $T_c \sim 3.7$ K. for Sample 3, together with substantially larger low-temperature superfluid densities compared with the sample discussed in the main text. In addition, $n_s$ displays a more complex temperature dependence, with no extended linear-in-$T$ behaviour as observed in the main sample.